\documentclass{llncs}
\usepackage{epsfig}

\newcommand{\dnaseq}[1]{\ensuremath{\mathtt{#1}}}
\newcommand{\dnaseqc}[1]{\ensuremath{\overline{\mathtt{#1}}}}
\newcommand{\ENC}{\ensuremath{\mathrm{enc}}}
\newcommand{\GENC}{\ensuremath{\mathrm{genc}}}
\newcommand{\etal}{{\it et al.}}
\newcommand{\RL}[1]{\textbf{\sffamily #1}}

\begin{document}

\title{DNA Self-Assembly For Constructing 3D Boxes}

\author{
Ming-Yang Kao\inst{1}\fnmsep\thanks{Supported in part by NSF Grants CCR-9531028 and EIA-0112934.  Part of this work was performed while this author was visiting the Department of Computer Science, Yale University, New Haven, CT 06520-8285, USA, \texttt{kao-ming-yang@cs.yale.edu}.}
and
Vijay Ramachandran\inst{2}\fnmsep\thanks{Supported by a 2001 National Defense Science and Engineering Graduate Fellowship.}
}

\institute{
Department of Computer Science, Northwestern University, Evanston, IL 60201, USA, \texttt{kao@cs.northwestern.edu}
\and
Department of Computer Science, Yale University, New Haven, CT 06520-8285, USA, \texttt{vijayr@cs.yale.edu}
}

\maketitle

\begin{abstract}
We propose a mathematical model of DNA self-assembly using 2D tiles to
form 3D nanostructures.  This is the first work to combine studies in
self-assembly and nanotechnology in 3D, just as Rothemund and Winfree
did in the 2D case.  Our model is a more precise superset of their Tile Assembly Model
that facilitates building scalable 3D molecules.
Under our model, we present algorithms to build a hollow cube,
which is intuitively one of the simplest 3D structures to
construct. We also introduce five basic measures of complexity to
analyze these algorithms. Our model and algorithmic techniques
are applicable to more complex 2D and 3D nanostructures.
\end{abstract}

\section{Introduction}\label{sec:intro}
DNA {\it nanotechnology} and DNA {\it self-assembly} are two 
related technologies with enormous potentials.

The goal of DNA nanotechnology is to construct small objects with high
precision. Seeman's visionary work \cite{see82} in 1982 pioneered the
molecular units used in self-assembly of such objects.  More than a
decade later, double-crossover (DX) molecules were proposed by Fu and
Seeman~\cite{fs93} and triple-crossover (TX) molecules by
LaBean~\etal~\cite{labean} as DNA
self-assembly building blocks.  Laboratory efforts have been
successful in generating interesting three-dimensional (3D) molecular structures,
including the small cube of Chen and Seeman \cite{cs91}.  However,
these are immutable and limited in size, mainly because their
fabrication is not based on a
mathematical model that can be extended as necessary.

In parallel to DNA nanotechnology, studies on self-assembly of DNA
tiles have focused on using local deterministic binding rules to
perform computations.  These rules are based on interactions
between exposed DNA sequences on individual tiles;
tiles assemble into a particular 1D or 2D structure when in solution,
encoding a computation. Winfree \cite{win96} formulated a
model for 2D computations using DX molecules.
Winfree~\etal~\cite{string} used 1D tiles for 1D computations and 2D
constructions with DX molecules. LaBean~\etal~\cite{lwr99} were the
first to compute with TX molecules.

Combining these two technologies, several researchers have
demonstrated the power of DNA self-assembly in nanostructure
fabrication.  Winfree \etal\ \cite{win98} investigated how to use
self-assembly of DX molecules to build 2D lattice DNA crystals.
Rothemund and Winfree~\cite{stoc2000} further proposed a mathematical
model and a complexity measure for building such 2D structures.

A natural extension of the seminal 2D results of
Winfree~\etal~\cite{win98} and Rothemund and Winfree~\cite{stoc2000}
would be the creation of 3D nanostructures using tiling.  To initiate
such an extension, this paper (1) proposes a general mathematical
model for constructing 3D structures from 2D tiles; (2) identifies a
set of biological and algorithmic issues basic to the implementation
of this model; and (3) provides basic computational concepts and
techniques to address these issues.  Under the model, the paper
focuses on the problem of constructing a hollow cube, which is
intuitively one of the simplest 3D structures to construct. We present
algorithms for the problem and analyze them in terms of five
basic measures of complexity.

There are three natural approaches to building a hollow cube. The
first approach uses 1D tiles to form 2D DX-type tiles as in \cite{string}, and
then uses these tiles to construct a cube.
Our paper does not fully investigate this possibility because of the
incovenient shape of these molecules (see Sect.~\ref{sec:tilemol}), but our algorithms
can be modified to accommodate these DX-type tiles.
The second approach builds a cube from genuine 2D
tiles, which is the focus of this paper.  The third approach is
perhaps the most natural: build a cube from genuine 3D tiles.  It
is not yet clear how such 3D tiles could be created; conceivably, the
cube of Chen and Seeman~\cite{cs91} may lead to tiles of this
form. This paper does not fully investigate this possibility, either,
because this approach is algorithmically straightforward
and similar to the 2D case.

\begin{figure}
\begin{center}
\scalebox{.75}{
\includegraphics{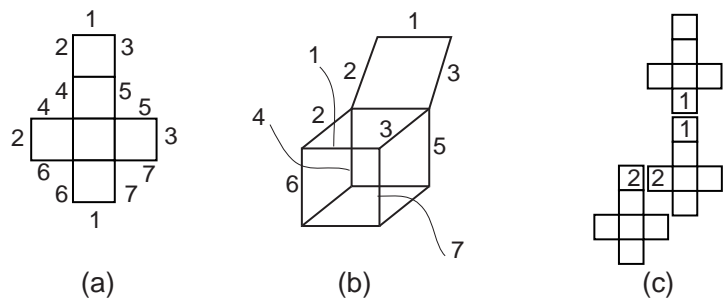}
}\end{center}
\caption{\label{fig:folding} {\bf (a)} 2D planar shape that will fold
into a box.  Each section is formed from many smaller 2D DNA tiles.
Edges with the same number have complementary sticky ends exposed so
they can hybridize.  {\bf (b)} Folding of the shape in (a) into a box.
Here, edges 4, 5, 6, and 7 have all hybridized.  Hybridization of
edges 2 and 3, whose two complements are now in close proximity, will
cause edge 1 to hybridize and form the complete box. {\bf (c)}
Multiple copies of the 2D shape in solution.  Copies of the shape can
interfere and attach infinitely without control as long as edges have
matching sticky ends.}
\end{figure}

The basic idea of our algorithms is to use 2D tiles to form a shape on
the plane that can fold into a box, as illustrated in
Fig.~\ref{fig:folding}(a)--(b).  We can easily synthesize
a set of tiles to create the intitial 2D shape.  
To overcome a negligible probability of success due to biochemical factors,
we must put many copies of these tiles into solution at once;
but we must then worry about multiple copies of the shape interfering
with each other, preventing folding, as in 
Figure~\ref{fig:folding}(c).

To avoid this problem, we introduce randomization, so that different
copies of the shape have unique sticky ends.
The growth of tiles into a complete structure must still be deterministic
(as it is based on Watson-Crick hybridization), but we randomize the
computation input --- the \emph{seed tiles} from which the rest of the shape assembles.
The edges then still relate to each other, but
depend on the random input that is different for each shape in solution.
If each input can form
with only low probability, interference with another copy of the shape
will be kept to a minimum.

This raises another important issue --- that of using self-assembly to
communicate information from one part of the shape to another.  Since
the edges must relate to each other and the random input, designing
local rules becomes nontrivial.  In this paper, we explore and
formalize patterns used in completing this task.  In addition, we
formalize biological steps that allow a specific subset of tiles to be
added in an isolated period of time, thus allowing better control of
growth.  We couple this with the use of temperature to improve the
probability of a successful construction.

The remainder of this paper is organized as follows.
Section~\ref{sec:model} describes the model of computation,
including notation for DNA tiles and definitions of complexity measures.
Section~\ref{sec:alg} describes the algorithms in detail, and 
Section~\ref{sec:openprobs} discusses future research possibilities.

\section{Model of Computation}\label{sec:model}

In this section we formally introduce our model of self-assembly,
the \emph{Generalized Tile Assembly Model}, on both the
mathematical and biological level.  It is an extension of the model
presented by Rothemund and Winfree in \cite{stoc2000}.

\subsection{Molecular Units of Self-Assembly}\label{sec:tilemol}

We begin with the biological foundation for our model.  We intend to
build 3D structures using the folding technique shown in Fig.~\ref{fig:folding}
and allow construction of all 2D structures possible with the Tile Assembly Model.

Our model relies on using the molecular building block of a \emph{DNA tile}.  Tiles can naturally
hybridize to form stable shapes of varying sizes, and the individual tiles can easily
be customized and replicated
(via synthesis and PCR before the procedure) for a specific algorithm.

DNA tiles are small nucleotides with exposed action sites (also known as sticky ends
of a DNA strand) consisting of a single-stranded sequence of base pairs.
When this sequence matches a complementary sequence on an action site of another tile,
the Watson-Crick hybridization property of DNA causes these two molecules to bind together,
forming a larger structure.  A tile can be synthesized in the laboratory to have specific sticky ends.
Different combinations of sticky ends on a tile essentially yield uniquely-shaped puzzle pieces.
The tiles will automatically hybridize when left in solution.

\begin{figure}
\begin{center}\scalebox{.6}{
\includegraphics{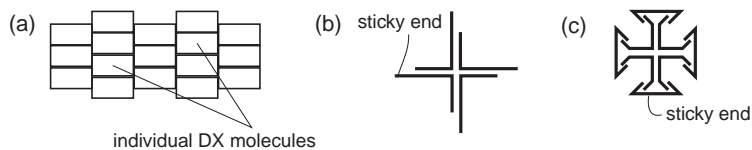}
}\end{center}
\caption{\label{fig:mol} {\bf (a)} 2D structure formed from DX molecules.  The left and right sides cannot
hybridize because they are aligned improperly;  the same is true for the top and bottom sides.
{\bf (b)} Branched-molecule DNA tile from \cite{see82}.  {\bf (c)} Snynthetic DNA tile derived from the structure of
tRNA.
}
\end{figure}

Most work in self-assembly uses DX and TX molecules for tiles, but the shape of these molecules causes a problem
for 3D construction.
Since the sticky ends are on diagonally opposite ends (see \cite{fs93} and \cite{labean}), these tiles form structures
with ragged edges when they hybridize, as in Figure~\ref{fig:mol}(a).  Our algorithms can easily be modified
to use these tiles by adjusting for proper alignment before folding into a box.

However, we propose a simpler alternative, which is using the branched molecules
of Seeman \cite{see82} or a variant derived from the structure
of tRNA.  These molecules, sketched in Figures~\ref{fig:mol}(b) and (c), are truly 2D with sticky
ends on four sides.  The structure is stable while the sticky ends are free-floating in solution --- so the
molecules have flexibility to align properly during folding.

Such molecules offer a natural motivation for modeling them using Wang's theory of tiling \cite{wang}, which
allows us to abstract construction using these molecules to a symbolic level.

\subsection{Symbolic Representation of Tiles}\label{sec:tilenot}

\begin{definition}A \emph{DNA sequence} of length $n$ is an ordered sequence of base pairs
$5'-\dnaseq{b}_{1}\dnaseq{b}_{2}\cdots\dnaseq{b}_{n}-3'$ where the sequence has a \emph{5-prime}
and \emph{3-prime} end, and $\dnaseq{b}_{i} \in \mathcal{B} = \{\dnaseq{A}, \dnaseq{T}, \dnaseq{C}, \dnaseq{G}\}$,
the set of base pairs.  We will assume that if the directions are not explicity written, the sequence
is written in the $5'\rightarrow 3'$ direction.\begin{enumerate}

\item The \emph{Watson-Crick complement} of sequence $\dnaseq{s} = 5'-\dnaseq{b}_{1}\dnaseq{b}_{2}\cdots\dnaseq{b}_{n}-3'$,
denoted \dnaseqc{s}, is the sequence\footnote{%
The assumption that sequences written without directions are given $5'\rightarrow 3'$ means that
the complement of $\dnaseq{b}_{1}\dnaseq{b}_{2}\cdots\dnaseq{b}_{n}$ is $\dnaseqc{b}_{n}\cdots\dnaseqc{b}_{2}\dnaseqc{b}_{1}$,
which is not standard convention but is technically correct.%
} $3'-\dnaseqc{b}_{1}\dnaseqc{b}_{2}\cdots\dnaseqc{b}_{n}-5'$, where
$\dnaseqc{A} = \dnaseq{T}$, $\dnaseqc{C} = \dnaseq{G}$.  Define $\overline{\dnaseqc{s}} = \dnaseq{s}$.

\item The \emph{concatenation}
of two sequences $\dnaseq{s} = \dnaseq{s}_{1}\cdots\dnaseq{s}_{n}$ and $\dnaseq{t} = \dnaseq{t}_{1}\cdots\dnaseq{t}_{m}$,
denoted $\dnaseq{s}\cdot\dnaseq{t}$, or simply \dnaseq{st}, is the sequence $\dnaseq{s}_{1}\cdots\dnaseq{s}_{n}\dnaseq{t}_{1}\cdots\dnaseq{t}_{m}$.

\item The \emph{subsequence from $i$ to $j$} of sequence $\dnaseq{s} = 5'-\dnaseq{b}_{1}\dnaseq{b}_{2}\cdots\dnaseq{b}_{n}-3'$,
denoted $\dnaseq{s}[i:j]$, is the sequence $5'-\dnaseq{b}_{i}\dnaseq{b}_{i+1}\cdots\dnaseq{b}_{j-1}\dnaseq{b}_{j}-3'$,
where $1 \leq i < j \leq n$.

\end{enumerate}
Given the above definitions, two DNA strands can \emph{hybridize} if they have complementary sequences.  Formally,
$\dnaseq{s} = \dnaseq{s}_{1}\cdots\dnaseq{s}_{n}$ and $\dnaseq{t} = \dnaseq{t}_{1}\cdots\dnaseq{t}_{m}$ can hybridize
if there exist integers $h_{s1}, h_{s2}, h_{t1}, h_{t2}$ such that $\dnaseq{s}[h_{s1}:h_{s2}] = \overline{\dnaseq{t}[h_{t1}:h_{t2}]}$.
We assume there are no \emph{misbindings}, that is, the above condition must be met exactly with no errors in base-pair binding.
\end{definition}

\begin{remark}
Note that \dnaseqc{st} (or \dnaseqc{(s\cdot t)}\,) $\neq$ \dnaseqc{s}\dnaseqc{t} (or $\dnaseqc{s}\cdot\dnaseqc{t}$); rather,
$\dnaseqc{st} = \dnaseqc{t}\cdot\dnaseqc{s}$.
\end{remark}

\begin{definition}
The \emph{threshold temperature} for a DNA sequence is a temperature $t$ in some fixed set $\mathcal{T}$
such that the sequence is unable to remain stably hybridized to its complement when the solution is at a temperature
higher than $t' \in (t-\epsilon, t+\epsilon)$ for $\epsilon > 0$.%
\footnote{A fixed set of threshold temperatures simplifies the model and corresponds to the temperature parameter
in \cite{stoc2000}.  To compensate we allow the actual threshold temperature to deviate slightly from the fixed point.} %
(Heating a solution generally denatures strands, so this definition has strong biological foundation.
The consequences and methodology of using temperature in designing DNA sequences for tiles is discussed
in \cite{words}.)  If \dnaseq{s} has a lower threshold temperature than \dnaseq{t},
we say \dnaseq{s} \emph{binds weaker} than \dnaseq{t}.
\end{definition}

As with most work in DNA computing, our model uses DNA sequences to encode information%
\footnote{Condon, Corn, and Marathe \cite{words} have done work on designing good DNA sequences for problems like this one.}%
 --- in our case, an identifier
specifying what kinds of matches are allowed between tiles on a given side.  Since there are no misbindings, these
identifiers map uniquely to DNA sequences present on the sides of tiles that can bind to each other.  Formally, we have
the following.

\begin{definition}
Let $\mathcal{S}$ be the set of \emph{symbols} used to represent the patterns on the sides of our tiles.  We assume
$\mathcal{S}$ is closed under complementation, that is, if $s \in \mathcal{S}$ then there exists some $s' \in \mathcal{S}$
such that $s' = \overline{s}$ where $\overline{s}$ is the \emph{complement} of $s$ (the purpose
of this will be clear below).  Let $\mathcal{W} \subset \bigcup_{i} \mathcal{B}^{i}$ be the set of DNA sequences called
\emph{DNA words} such that the words do not interfere with each other or themselves (i.e., bind inappropriately).
We then define the injective map $\ENC:\,\mathcal{S}\rightarrow\mathcal{W}$ that is the \emph{encoding} of a symbol into a DNA word.
This map obeys \emph{complementation}: $\ENC\left(\overline{s}\right) = \overline{\ENC(s)}$.
\end{definition}

\begin{definition}
A \emph{DNA tile} is a 4-tuple of symbols $T = (s_{N}, s_{E}, s_{S}, s_{W})$ such that
$s_{i} \in \mathcal{S}$ and $\ENC(s_{i})$ is the exposed DNA sequence at the north, east, south, or west action site
of the tile, for $i = N, E, S, W$.  Given two tiles $T_{1}$ and $T_{2}$, they will bind if two sides have complementary
symbols.  Properties of hybridization, including threshold temperature, carry over to the hybridization of tiles.
We make a stronger no-misbinding assumption for tiles, requiring that the sticky ends on the tiles match exactly and fully.
\end{definition}

At this stage, our model exactly matches that of Rothemund and Winfree in \cite{stoc2000}, except that our tiles
can ``rotate''; that is, $(s_{N}, s_{E}, s_{S}, s_{W}) = (s_{E}, s_{S}, s_{W}, s_{N})$.  This corresponds
more closely to tile structure.  The model, at this point, could require many symbols to express different tile types,
and possibly an exponential number of DNA words. Ideally, we would like to arbitrarily extend the
symbolic or informational content of each side of a tile.  Therefore we make the following generalization.
\begin{definition}
Let $\Sigma$ be a set of symbols closed under complementation, and let $\Omega$ be a set of corresponding DNA words.
A \emph{$k$-level generalization} of the model defines a map $g:\,\Sigma^{k}\rightarrow\mathcal{S}$ and a
corresponding encoding $\GENC:\,\Sigma^{k}\rightarrow\mathcal{W}$, where $\GENC(\sigma) = \ENC(g(\sigma))$ for $\sigma \in \Sigma^{k}$
such that an \emph{abstract tile definition}, which is a 4-tuple of $k$-tuples of symbols in $\Sigma$, is equivalent
to a DNA tile.

We define complementation for a $k$-tuple in $\Sigma^{k}$ as follows:  let $\overline{\sigma} = \overline{(\sigma_{1}, \ldots, \sigma_{k})}$
be $(\overline{\sigma_{1}}, \ldots, \overline{\sigma_{k}})$ so that $\GENC(\overline{\sigma}) = \ENC(g(\overline{\sigma}))
= \ENC(\overline{g(\sigma)}) = \overline{\ENC(g(\sigma))}$.  This makes the hybridization condition equivalent to
having complementary symbols in $k$-tuples for sides that will bind.
\end{definition}
The definition is purposefully broad in order to allow different algorithms to define the encoding based on the number
of words and tiles needed.  A $1$-level generalization with $\Sigma = \mathcal{S}$ and $\Omega = \mathcal{W}$
where $g(s) = s$ and $\GENC = \ENC$ is the original Rothemund-Winfree Tile Assembly Model.  In this paper, we use the following model.
\begin{definition}
The \emph{concatenation generalization} is a $k$-level generalization where $\mathcal{W} \subset \Omega^{k}$
and $g$ maps every combination of symbols in $\Sigma^{k}$ to a unique symbol in $\mathcal{S}$.  Partition $\mathcal{S}$
into $\mathcal{S}'$ and $\overline{\mathcal{S}'}$ such that each set contains the complement symbols of the other, and
$\mathcal{S}' \cap \overline{\mathcal{S}'} = \emptyset$.%
\footnote{The map \ENC\ as defined will be one-to-one, and so is $g$, and so this can be done since the DNA sequence corresponding
to any symbol has a unique complement, and therefore a unique complement symbol.}
For $\sigma \in \mathcal{S}'$, define $\GENC(g^{-1}(\sigma)) = \ENC(\sigma) = \omega_{1}\omega_{2}\cdots\omega_{k}$,
where $\ENC(\sigma_{i}) = \omega_{i} \in \Omega$ and $g^{-1}(\sigma) = (\sigma_{1}, \sigma_{2}, \ldots, \sigma_{k})$.
Then for $\overline{\sigma} \in \overline{\mathcal{S}'}$, let $\ENC(\overline{\sigma}) =
\overline{\omega_{k}}\cdot\overline{\omega_{k-1}}\cdots\overline{\omega_{1}}$, so that $\GENC(g^{-1}(\overline{\sigma})) =
\overline{\GENC(g^{-1}(\sigma))}$.
\end{definition}
In other words, the concatenation generalization model is a straightforward extension of the tile model where each
side of a tile corresponds to a $k$-tuple of symbols, where the DNA sequence at the corresponding action site is simply
the concatenation of the encodings of the individual symbols.%
\footnote{Potentially, the concatenation model could cause interference among tiles.  If we maintain no misbindings,
however, our model removes this from analysis.  In addition, it is theoretically possible to design DNA words
so interference does not occur, depending on the algorithm.}
Using this simple model, we can reduce the number of
DNA words needed to $|\Sigma|$ from $|\Sigma|^{k}$, and create simpler descriptions of our tiles.

\subsection{Algorithmic Procedures}\label{sec:procs}

With the above models for tiles, we now discuss procedures for growing larger structures.

We follow Rothemund and Winfree \cite{stoc2000} and Markov \cite{crystal} and use the common
self-assembly assumption that a structure begins with a seed tile and grows, at each timestep, by hybridization
with another free-floating tile.\footnote{In reality, multiple tiles can hybridize at once and structures consisting
of more than one tile can hybridize to each other, but we lose no generality with the Markov assumption.}
The new tile hybridizes at a given position following one of two types of rules:\begin{description}
\item[Deterministic] Given the surrounding tiles at that position, only one tile type, with specific sticky ends on
the non-binding sides, can fit.

\item[Randomized]  Multiple tile types (with different sticky ends on the non-binding sides) could fit the position
given the tiles present; a new action site is created with probability proportional to the concentration
of its tile type in solution.
\end{description}

Therefore, to grow a structure, an algorithm repeats \emph{steps} until the structure is complete:
add tiles to solution; wait for them to adhere to the growing structure;
optionally removes excess tiles from solution by ``washing them away.''  Cycling temperature during these
steps to prevent or induce binding (based on threshold temperatures) can be done while waiting for hybridization,
and is called \emph{temperature-sensitive binding}.

\subsection{Complexity}\label{sec:complx}

We consider five basic methods of analyzing algorithms using our model.\begin{description}
\item[Time complexity] Each algorithm is a sequence of self-assembly steps
described above, thus the natural measure of
\emph{time complexity} in our model is the number of
steps required (which describes laboratory time).

\item[Space complexity] The number of distinct physical tile types
(not the actual number of molecules produced) is
\emph{space complexity}.  Introduced by \cite{stoc2000}, this describes
the amount of unique DNA synthesis necessary.

\item[Alphabet size] The number of DNA words, or $|\Omega|$ or $|\mathcal{W}|$, has a rough laboratory
limit \cite{words}, and so the size of the symbol set used ($|\Sigma|$ or $|\mathcal{S}|$), which corresponds directly to the number of words,
has practical significance.

\item[Generalization level] The generalization level is
the amount of information on a side of a tile.  This is related to the length of the sticky ends (and
thus has biological consequences) and the number of actual DNA words (via $|\mathcal{S}|$).

\item[Probability of misformation]
\emph{Misformed structures} contain tiles that are not bound properly on all sides.  Assuming the Markov model,
consider adding tile $T$ to a partial structure $S$.  If complete hybridization requires binding on two sides,
but $T$ manages to hybridize only on one side (while the
other action site does not match), $S+T$ has a \emph{misformation}.  We quantify this probability with the following.
\begin{definition}
Let the \emph{success probability at step $t$} be the probability that a free-floating tile in solution
binds at all possible sides to a partial structure at a given spot.  (Step $t$ is the addition of a tile
to that spot on structure $S_{t}$, resulting in $S_{t+1}$.)  This is
\[\mathrm{Pr}(S_{t+1}\mbox{ is correct}\,|\, S_{t}\mbox{ is correct}) = \frac{N_{\mbox{\small correct}}}{N_{\mbox{\small all}}}\enspace,\]
where $N_{\mbox{\small correct}}$ is the number of tile types that can correctly bind, while $N_{\mbox{\small all}}$ is the number
of tile types in solution that could bind, possibly even incompletely.
Call this $q_{t}$.  Then the \emph{misformation probability} at step $t$ is $p_{t} = 1-q_{t}$.  If the algorithm has $k$
additions, then the \emph{misformation probability} for the algorithm is $1-q_{0}q_{1}\cdots q_{k-1}$.
Then an algorithm is \emph{misformation-proof} if its misformation probability at every step is zero, yielding
a zero total probability of misformation.\end{definition}
\end{description}

\section{Hollow Cube Algorithms}\label{sec:alg}

In this section, we examine algorithms designed to use our model to build a 3D hollow cube
using the folding technique shown in Fig.~\ref{fig:folding}.  Let the length of a side of the cube, $n$,
be the input to the algorithms.  We present the most interesting algorithm in detail and discuss
some others briefly.

\subsection{Overview}\label{sec:alg-o}
\begin{figure}
\begin{center}\scalebox{0.6}{
\includegraphics{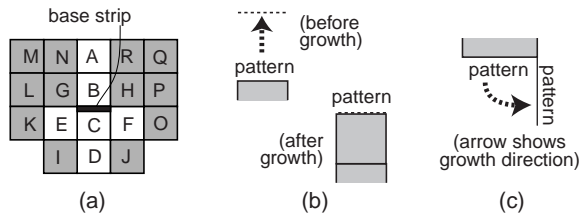}
}\end{center}
\caption{\label{fig:rbr} {\bf (a)} Regions of the 2D planar shape.
{\bf (b)} A straight-copy pattern.  {\bf (c)} A turn-copy pattern.
}
\end{figure}

Figure~\ref{fig:rbr}(a) illustrates the planar shape our algorithms construct.
We will reference the labels and shading of regions in the figure
during our discussion.

As stated earlier in Sect.~\ref{sec:intro}, we must make each shape unique so different partial structures
in solution do not bind to and interfere with each other.  Once we have a unique seed structure, we can then
use self-assembly with basic rules to make the edges of the shape correspond so folding will occur.

There are three basic self-assembly \emph{patterns} used to construct different
parts of the shape.\begin{description}

\item[Random assembly] Implements a random rule (see Sect.~\ref{sec:procs}).  Formally,
add all tiles in a set of distinct tiles $R$ in equal concentrations so each
could potentially hybridize completely at a given position.  Thus the information at that position
is completely random.  The tiles differ by a component of their exposed $k$-tuples, assuming
a $k$-level generalization.

\item[Straight copy] See Figure~\ref{fig:rbr}(b).  Tiles are added to copy the pattern
along one end of a region through to a parallel end of an adjacent region being constructed.
This rule is deterministic.

\item[Turn copy] See Figure~\ref{fig:rbr}(c).  Tiles are added to copy the pattern along
one end of a region to a perpendicular end of an adjacent region being constructed.  Counters
will be required to position the tiles appropriately to complete this deterministic rule.
\end{description}
The algorithm begins by assembling a random pattern string that will be copied to the top
and bottom of the box.  Then random patterns are added for the remaining edges of the box,
and these are copied to the corresponding edges accordingly.  (Refer to Fig.~\ref{fig:folding}
for corresponding edges.)  Finally, the shape that will fold is cut out by raising the temperature,
assuming that the bonds between tiles along the region-borderline have weak threshold temperatures.
The regions that are shaded in Fig.~\ref{fig:rbr}(a) are cut away.

\subsection{Notation}\label{sec:symbols}
All of our algorithms will use a $3$-level concatenation generalization model; thus a tile is a 4-tuple of triplets
that we write $T_{N} \times T_{S} \times T_{W} \times T_{E}$, with each $T_{i} = (\sigma_{1}, \sigma_{2}, \sigma_{3})$.
We change the order of the tuple to
more easily identify tiles that will bind, since most binding will be north-south or west-east.
(This decision is arbitrary and purely notational.)  We
assume all tiles are oriented so that the directions are clear.

We define the set $\Pi \subset \Sigma$ to be the ``random patterns'' $\pi_{1}, \pi_{2}, \ldots, \pi_{p}$
used as components of exposed triplets for tiles used in random assembly;  their use will become clear
when we discuss implementation of random assembly below.

We use counters to control the growth of 
of our planar shape.  This concept has been well explored in \cite{stoc2000} and earlier papers.
Each tile can be assigned a position in the plane denoted by a horizontal and vertical coordinate.
We create symbols for position counters
and then allow tiles to hybridize if the positions match, creating a mechanism for algorithms to place
tiles in absolute or relative positions. Let $H(i)$ and $V(j)$ be the symbols
denoting horizontal position $i$ and
vertical position $j$, respectively.

\subsection{Row-by-row Algorithm}\label{sec:algrbr}
The row-by-row algorithm sacrifices time complexity to reduce
space complexity and alphabet size.  In addition,
by using more steps and temperature-sensitive binding, the algorithm eliminates the possibility of
misformations.

\subsubsection{Implementation of Random Assembly}

The base strip shown in Fig.~\ref{fig:rbr}(a) is created via random assembly and represents the unique seed structure
for the shape.  The pattern assembled on the strip will be copied to other edges.  The strip will have length $n$
as it is an edge of the cube, and so we use horizontal counters to control growth.  We add the following tiles
to solution:

{\scriptsize\begin{eqnarray}
& \left(\beta, H(0), V(0)\right) \times \left(\overline{\beta},\overline{H(0)},\overline{V(-1)}\right) \times \left(\rho',H(0),V(0)\right) \times\left(\overline{\kappa_{1}},\overline{H(1)},\overline{V(0)}\right) & \label{eqn:rbr-end1}\\
& \left(\beta, H(0), V(0)\right) \times \left(\overline{\beta},\overline{H(0)},\overline{V(-1)}\right) \times \left(\kappa_{1},H(n-1),V(0)\right) \times\left(\rho',H(0),V(0)\right) & \label{eqn:rbr-end2} \\
& \left(\alpha, \kappa_{2}, V(0)\right) \times \left(\overline{\alpha},\overline{\kappa_{2}},\overline{V(-1)}\right) \times \left(\kappa_{1},H(1),V(0)\right) \times\left(\overline{\kappa_{1}},\overline{H(2)},\overline{V(0)}\right) & \label{eqn:rbr-end3} \\
& \left(\alpha, \kappa_{2}, V(0)\right) \times \left(\overline{\alpha},\overline{\kappa_{2}},\overline{V(-1)}\right) \times \left(\kappa_{1},H(n-2),V(0)\right) \times\left(\overline{\kappa_{1}},\overline{H(n-1)},\overline{V(0)}\right) & \label{eqn:rbr-end4} \\
& \left(\pi_{k}, \kappa_{2}, V(0)\right) \times \left(\overline{\pi_{k}},\overline{\kappa_{2}},\overline{V(-1)}\right) \times \left(\kappa_{1},H(i),V(0)\right) \times\left(\overline{\kappa_{1}},\overline{H(i+1)},\overline{V(0)}\right) \label{eqn:rbr-end5} & 
\end{eqnarray}}

\noindent where the tiles in (\ref{eqn:rbr-end5}) vary over all $i$, $2 \leq i \leq n-2$, and all $k$, $1 \leq k \leq |\Pi|$.
Since the ends, given by (\ref{eqn:rbr-end1}) and (\ref{eqn:rbr-end2}), the tiles bordering the ends, given by (\ref{eqn:rbr-end3}) and (\ref{eqn:rbr-end4}), and the pattern
tiles in the middle of the strip, given by (\ref{eqn:rbr-end5}), all have appropriate horizontal counter markings $H(i)$
for every position $0, \ldots, n-1$, every base strip must have the appropriate ends and length $n$.  Note that the
pattern on the exposed north and south side of tiles in position $2, \ldots, n-2$ given by (\ref{eqn:rbr-end5}),
is completely random as any one of the tiles containing some $\pi_{i}$ could hybridize there.  Thus, this shape now
has a unique identifier, namely, the sequence of patterns $\{\pi_{i}\}$ along the base strip;  the probability of
any sequence expressed on a given shape is $|\Pi|^{n-2}$.

\subsubsection{Implementation of Straight Copy}

We must complete a straight-copy from the base strip through regions \RL{A},\RL{B},\RL{C}, and \RL{D}, in addition
to straight-copy patterns required elsewhere.  This is done using $2n+1$ steps as the tiles for \RL{A} and \RL{C} are done in parallel (the same for \RL{B} and \RL{D}).
In the following example for a straight-copy through \RL{A}-\RL{D}, assume
that the triplet constant $X = \left(\varphi_{1},\varphi_{2},\varphi_{3}\right)$.  The following tiles are added for
each step $i$, $1 \leq i \leq 2n-2$:

{\scriptsize\begin{eqnarray}
& \left(\alpha,\kappa_{2},V(i)\right) \times \left(\overline{\alpha},\overline{\kappa_{2}},\overline{V(i-1)}\right) \times \left(\sigma_{1},\sigma_{2},V(i)\right) \times \overline{X} & \label{eqn:rbr-intaupl} \\
& \left(\alpha,\kappa_{2},V(i)\right) \times \left(\overline{\alpha},\overline{\kappa_{2}},\overline{V(i-1)}\right) \times X \times \left(\sigma_{1},\sigma_{2},V(i)\right) & \label{eqn:rbr-intaupr} \\
& \left(\alpha,\kappa_{2},V(-i)\right) \times \left(\overline{\alpha},\overline{\kappa_{2}},\overline{V(-i-1)}\right) \times \left(\sigma_{1},\sigma_{2},V(-i)\right) \times \overline{X} & \label{eqn:rbr-intadnl} \\
& \left(\alpha,\kappa_{2},V(-i)\right) \times \left(\overline{\alpha},\overline{\kappa_{2}},\overline{V(-i-1)}\right) \times X \times \left(\sigma_{1},\sigma_{2},V(-i)\right) & \label{eqn:rbr-intadnr} \\
& \left(\pi_{k},\kappa_{2},V(i)\right) \times \left(\overline{\pi_{k}},\overline{\kappa_{2}},\overline{V(i-1)}\right) \times X \times \overline{X} & \label{eqn:rbr-intpup} \\
& \left(\pi_{k},\kappa_{2},V(-i)\right) \times \left(\overline{\pi_{k}},\overline{\kappa_{2}},\overline{V(-i-1)}\right) \times X \times \overline{X} & \label{eqn:rbr-intpdn}
\end{eqnarray}}

\noindent Tiles versions exist for all patterns in $\Pi$ when $\pi_{k}$ is present in the tile description above.
We assume that $X$ binds weaker than $\kappa_{2}$, so cycling the temperature ensures the tiles are attached
on the $\kappa_{2}$ side.

The tiles are added sequentially to prevent misformations.  It is evident
that the borders of these regions, given by (\ref{eqn:rbr-intaupl})--(\ref{eqn:rbr-intadnr}) have symbols on the exposed sides
to hybridize with a strip of tiles that will form a folding edge of the box, and the middle tiles copy the random pattern to
another location.

After $2n-2$ of these steps, the middle regions are complete, except for two rows on top
(rows $2n-1$ and $2n$) and one row on the bottom ($2n-1$).  The order of these rows is important, as we add
an extra set of tiles to prevent the top and bottom from hybridizing before the folding is complete.

\subsubsection{Implementation of Turn Copy}

The turn-copy step, for example, copying the bottom edge of \RL{E} to the left edge of \RL{D} through \RL{I} so the shape can fold,
is done using vertical and horizontal counters, which essentially places a tile in a specific spot.  Therefore we can add
all the tiles at once to complete the region without possibility of misformation.
For the above example region, we would add the following tiles.
Let $i$ and $j$ vary such that
$-n \leq i \leq -1$ and $-2n+1 \leq j \leq -n$.  For all $i,j$, add:

{\scriptsize
\begin{equation}\label{eqn:rbr-turneq} i=j+(n-1),\;\left(\pi_{k},H(i),V(j)\right) \times \left(\overline{\kappa_{3}},\overline{H(i)},\overline{V(j-1)}\right) \times \left(\kappa_{3},H(i),V(j)\right) \times \left(\overline{\pi_{k}},\overline{H(i-1)},\overline{V(j)}\right) \end{equation}
\begin{equation}\label{eqn:rbr-turnl} i<j+(n-1),\;\left(\pi_{k},H(i),V(j)\right) \times \left(\overline{\pi_{k}},\overline{H(i)},\overline{V(j-1)}\right) \times \left(\kappa_{3},H(i),V(j)\right) \times \left(\overline{\kappa_{3}},\overline{H(i-1)},\overline{V(j)}\right) \end{equation}
\begin{equation}\label{eqn:rbr-turng} i>j+(n-1),\;\left(\kappa_{3},H(i),V(j)\right) \times \left(\overline{\kappa_{3}},\overline{H(i)},\overline{V(j-1)}\right) \times \left(\pi_{k},H(i),V(j)\right) \times \left(\overline{\pi_{k}},\overline{H(i-1)},\overline{V(j)}\right) \end{equation}
}

\noindent The above copies the pattern through \RL{I}, and the following adds the left edge to \RL{D}, where $-2n+1 \leq i \leq -n-1$:

{\scriptsize\begin{eqnarray*}
& \left(\beta,H(0),V(-n)\right) \times \left(\overline{\kappa_{3}},\overline{H(0)},\overline{V(-n-1)}\right) \times \left(\pi_{k},H(0),V(-n)\right) \times \left(\overline{\sigma_{1}},\overline{\sigma_{2}},\overline{V(-n)}\right) & \\
& \left(\kappa_{3},H(0),V(i)\right) \times \left(\overline{\kappa_{3}},\overline{H(0)},\overline{V(i-1)}\right) \times \left(\pi_{k},H(0),V(i)\right) \times \left(\overline{\sigma_{1}},\overline{\sigma_{2}},\overline{V(i)}\right) &
\end{eqnarray*}}

\noindent As an extra precaution we can set $\kappa_{3}$ binding to be weaker than $\pi_{k}$, and
cycle the temperature several times.  In addition, we force the encoding of horizontal and vertical counters
at the edges where the folding occurs to be the same, so that the sticky ends are in fact complementary.

\subsubsection{Summary}  After random assembly of the base strip, we use a row-by-row straight copy using a one-dimensional
counter through regions \RL{A}-\RL{D} to copy the base strip's pattern.  We then use straight copy to fill in the
bodies of regions \RL{E} and \RL{F} and add the edges using random assembly, as these will correspond to other portions
of the shape.  We then use a turn copy through \RL{G}-\RL{J} to make those edges correspond.  Finally, we do a sequence
of straight and turn copies from \RL{E} and \RL{F} through \RL{K}-\RL{N} and \RL{O}-\RL{R} to complete the shape.
We then raise the temperature to cut away the shaded regions.

\subsubsection{Analysis of the Row-by-row Algorithm}

\begin{theorem}$|\Sigma| = 8n + |\Pi| + O(1)$.\end{theorem}
\begin{proof}
The symbol set $\mathcal{S}$ consists of the following: pattern symbols in $\Pi$;
$4n$ horizontal and $4n$ vertical counters for positions throughout the shape;
and some specialized symbols, not dependent on the inputs.
\qed\end{proof}

\begin{theorem}The algorithm has
has time complexity approximately $5n$.\end{theorem}
\begin{proof}
The algorithm consists of the following steps:
1 step to build the base strip;
$2n+1$ steps to complete \RL{A}-\RL{D};
$n$ steps to grow \RL{E} and \RL{F};
1 step to add random patterns to the leaf borders;
1 step for the first set of turn-copy steps (\RL{G},\RL{I},\RL{H},\RL{J});
and $2n+2$ steps to complete regions \RL{K}-\RL{N} and \RL{O}-\RL{R}.
The time complexity for the basic version, therefore, is approximately $5n$ steps.
\qed\end{proof}

\begin{theorem}The space complexity of the row-by-row algorithm is approximately $6|\Pi|n^{2}+10|\Pi|n+4|\Pi|+8n$ tiles.\end{theorem}
\begin{proof}Clear from counting the number of tiles required,
including all the necessary variations of the same tile for position and pattern.
\qed\end{proof}

\begin{theorem}The number of distinct temperatures required is 3.\end{theorem}
\begin{proof}One temperature is required for detaching the excess portions of the 2D shape, and the highest temperature
level is for the main portion of the box that remains intact.  Adding tiles in rows involves cycling the temperature
to prevent misformations, by ensuring the ``random-patterned'' side of the tiles bind.  This requires a third temperature
between the two discussed where potential misformations denature.
\qed\end{proof}

\begin{theorem}The misformation probability of row-by-row is 0.\end{theorem}
\begin{proof}The random assembly steps grow the shape in one direction using a one-dimensional counter,
and only the correct tiles are marked with the appropriate counters.
Thus the probability for these steps is zero.
The turn-copy steps use both
vertical and horizontal counters defining a two-dimensional position that bind weaker than the random patterns.
As these are the only three words on tiles in this pattern, these also have a misformation probability
of zero.  Finally, because the straight-copy steps are performed in a sequence and all the potential spots
for misformation are sealed with tiles one row at a time, the probability of misformation here is also zero.
Thus the misformation probability of the algorithm is 0.
\qed\end{proof}

\subsection{All-together Algorithm}
The row-by-row algorithm does not take full advantage of the parallelism of self-assembly.
Instead of adding one row after another, we can synthesize tiles with both horizontal and vertical counters,
so each tile type can fit at only one absolute position.  Then
we can begin with virtually all our tiles in solution and allow the assembly to proceed in constant time.

Random assembly is performed as above, with special one-dimensional counters to control length.  However, these
strips must be assembled separately and independently to prevent interference.  The sequences exposed opposite the
random patterns will dictate where these strips can later hybridize on the larger shape, and the random patterns
will cause growth that makes the shape unique, as needed.
The turn copy implementation is unchanged, as counters are already present.
The straight copy, rather than using a ``constant'' symbol,
must instead use counters for growth in both directions.

Although this algorithm is straightforward, with all the tiles in solution at once, we might get incomplete
partial structures.  Specifically, tiles can bind on sides that have counters but do not have any $\pi_{i}$ component,
thus creating shapes with no information allowing them to fold.  In this case, at any of the $O(n^{2})$ positions,
a misformation could occur if the counters bind but the patterns do not, and so the misformation probability
of all-together is $1-|\Pi|^{-n^{2}}$.  We can prevent this by requiring the counters to bind weaker than the
$\pi_{i}$ and cycle the temperature (requiring 3, rather than 2, distinct temperature levels),
giving a misformation probability of 0, but practically, this has little meaning
since we may still get incomplete, partial structures and may have to wait a long time for proper hybridization
to occur.

Besides this drawback, this algorithm performs well by other complexity measures.  Since counters are necessary
for the row-by-row algorithm, the alphabet size is not greatly increased.  The algorithm has space complexity
$O(n^{2}|\Pi|)$, essentially because a different tile type is created for every position and every pattern
that could be present there.

\subsection{Other Algorithms}
We have considered several other algorithms using the same $3$-level generalization model, experimenting
with different uses or omissions of counters and other implementations of the self-assembly patterns.

The \emph{by-region} algorithm keeps space complexity and time complexity low by removing counters
and controlling growth in only certain rows and columns of a region.  The probability of misformation is increased.

The \emph{border-first} algorithm has the highest misformation probability, although has a very low time and
space complexity.  In this algorithm, the frame of regions are constructed first, when possible, and then
``filler tiles'' are added in to strengthen the structure later.  There are potential stability problems
during shape growth.

Finally, another possibility is to build the six faces separately and allow them to hybridize, but with this method
there is little control over how the final shape actually forms in solution.

\section{Conclusion}\label{sec:openprobs}

Our paper introduces a precise extension to the Tile Assembly Model \cite{stoc2000} that allows
greater information content per tile and scalability to three dimensions.  The model better formalizes
the abstraction of DNA tiles to symbols, introduces five complexity measures to analyze algorithms,
and is the first to extend nanostructure fabrication to three dimensions.

In addition, our paper opens up wide-ranging avenues of research.

First of all, it may be possible to encode information on tiles more succintly than our algorithms
do to accomplish the copy patterns discussed.  The existence of
good $2$-level or $1$-level generalization algorithms is unknown.

Algorithms to form other 3D structures, having various applications in biology and computation,
can be studied.  More work also must be done to quantify the probabilities
specified in the paper (possibly including a free-energy analysis of tile binding).

Finally, there remain some
important biological issues.  In particular, design of a strong tile
suitable for our method of computation and design of a 3D
building block are two important steps to increasing the feasibility
of 3D self-assembly.  The use of temperature may be further refined and exploited
to improve some complexity results and the number of steps needed in
the lab.

\bibliographystyle{abbrv}
\bibliography{dna_3d}

\end{document}